\documentclass{ws-ijmpa}

\begin{document}

\markboth{O.M. Lecian, G. Montani}
{Fundamental Symmetries of the extended Spacetime}

%
\catchline{}{}{}{}{}
%

\title{FUNDAMENTAL SYMMETRIES OF THE EXTENDED SPACETIME 
}

\author{ORCHIDEA MARIA LECIAN
}

\address{Dipartimento di Fisica and ICRA, Sapienza, University of Rome,\\Piazzale Aldo Moro, 5- 00185 Roma, Italy
\\
lecian@icra\_it}

\author{GIOVANNI MONTANI}

\address{Dipartimento di Fisica and ICRA, Sapienza, University of Rome,\\Piazzale Aldo Moro, 5- 00185 Roma, Italy\\
ENEA -- C.R. Frascati (Department F.P.N.),\\Via 
Enrico Fermi, 45- 00044, Frascati (Roma), Italy\\
ICRANet -- C. C. Pescara,\\Piazzale della 
Repubblica, 10- 65100, Pescara, Italy\\
montani@icra\_it}

\maketitle

\begin{history}
\received{Day Month Year}
\revised{Day Month Year}
\end{history}

\begin{abstract}
On the basis of Fourier duality and Stone-von Neumann theorem, we will examine polymer-quantization techniques and modified uncertainty relations as possible 1-extraD compactification schemes for a phenomenological truncation of the extraD tower.

\end{abstract}

\ccode{PACS numbers: 04.50.Cd, 04.60.Nc, 02.30.Px}

\section{Basic Statements}
In the simplest model, i.e., a scalar field in a 5-dimensional (5D) spacetime, described as the direct product of a 4D manifold plus a ring, $M^5=M^4\otimes S^1$, the Kaluza-Klein (KK) tower is defined as
\begin{equation}\label{infinite}
\Psi^{5}(x^{\rho},x^{5})=\sum_{-\infty}^{+\infty}\psi_{m}(x^{\rho})e^{ix^{5}m/L}, \quad L\equiv 2\pi R,
\end{equation}
that is an infinite sum of Fourier harmonics\footnote{Because of the periodic (boundary) condition on the modes of the tower, $\psi(x^{5})=\psi(x^{5}+L)$, i.e., of the identification of the points $0\leftrightarrow 2\pi R$, $\psi(x_{5})$ is defined on $S_{1}/\mathbb{Z}\sim\mathbb{R}$.}, labeled by $m$\\
The scalar-field wavefunction obeys the Klein-Gordon equation 
and its expression in the momentum representation reads $\tilde{\psi}^{m}(P_{5})=\delta(P_{5}-m/L)$. From \ref{infinite}, it is easy to understand the the structure of the extraD geometry can be described by means of the extraD projection of physical objects, and truncating the expansion would correspond to modifying the underlying geometry and algebra\cite{dufdo}. The most dramatic truncation is considering $m=0$ only\cite{dol84}. Otherwise, if $N<\infty$, $N$ is the maximum number of particles allowed in the tower, which corresponds to possible $P_{5}$ discrete values. A possible strategy is studying a finite set of delta-approximating functions, whose finite sum should reproduce the periodicity on the extraD coordinate in the momentum space, and their FT's\footnote{For instance, Bessel or Airy functions approximate the delta function in the vicinity of the peak, but exhibit a behavior than cannot fit our purposes outside the considered region. Finite-width peaked oscillating functions, such as $\tilde{\psi}_{N}(P_{5}^{m})=\frac{1}{2\pi}\frac{\sin \left(N(P_{5}-m/L)\right)}{P_{5}-m/L}$ are peaked around the desired value, but then exhibit secondary peaks around all the other values, and, for this reason, do not allow for any proper definition of scalar product, even in the momentum representation. Finite-width peaked functions,
such as $\tilde{\psi}_{N}(P_{5}^{m})= \sqrt{\frac{N}{\pi}}e^{N(P_{5}-m/L)^{2}}$, which, because of the width finiteness, lead to ill defined objects.}. Nonetheless, the possibility of introducing a non-trivial measure and modified operators\cite{gra06} for these Hilbert spaces naturally suggests us to abandon these attempts and to turn at other models, where the Hilbert space is endowed with a non-trivial measure. Two alternative strategies will be addressed. We will first consider truncated wavefunction as a quasi-periodic function, projected on a finite set of Fourier modes: for this purpose, we will analyze different representations of the standard operator algebra for the canonical commutation relations of the extraD operators $\hat{x}_{5}$ and $\hat{P}_{5}$. We will then go the other way round and establish generalized commutation relations. A phenomenological comparison\cite{dem94} of the two scenarios is however not possible, because the Stone-von Neumann theorem cannot apply for discontinuous operators (in the first case) and for non-unitary transformations (in the second case). 
\section{Modified Operator Algebra: Polymer Representation}
The introduction of a non-trivial measure suggests us to hypothesize that we are dealing with a representation of the Weyl algebra of the operators $\{x_{5},P_{5}\}$, whose Schroedinger representation does not admit self-adjoint operators unless an extra term is appended, i.e., whose states belong to an Hilbert space equipped with a non-trivial measure, $H=L^{2}(\mathbb{R}, d\mu)$.\\
The truncated version of (\ref{infinite}) can be interpreted as consisting of true plane waves, $e^{ix^{5}m/L}$, and almost-periodic functions of $x^{5}$, $\Psi^{5}(x^{\rho},x^{5})$. This way, it is interesting to analyze the polymer representation\cite{cor07} of such a problem. In the polymer representation, a different (Weyl) algebra for the operators $\{ \hat{x}_{5},\hat{P}_{5}\}$ can be found, based on the implementation of the phase space of the operators $\{x_{5},P_{5}\}$ to a complex manifold, with a complex structure $J_{N}$ such that
\begin{equation}\label{complex}
J_{N}\quad :\quad (x_{5},P_{5})\rightarrow \left(-\frac{P_{5}}{N^2},x_{5}N^2\right),
\end{equation}
which depends on the number of particles explicitly. Starting from the canonical commutation relation $[x_{5},P_{5}]=i$, one takes the exponentiated version of the operators,
\begin{equation}
U(a)=e^{iax_{5}},\quad V(b)=e^{ibP_{5}},
\end{equation}
where $a$ and $b$ are two parameters, acting on the set of wave functions $\{ \phi \}$, which are mapped to the Schroedinger representation via an isometric isomorphism. We then choose a discrete momentum variable, so that the operator $\hat{x}_{5}$ does not exist in the limit $N\rightarrow\infty$, while its exponentiated version still does. This way, the representation of the algebra will be inequivalent to the usual one, because the Stone-von Neumann uniqueness theorem does not apply in this case. As a result, if $P_{5}$ is discrete, states decompose as $|\psi>=\sum_{0}^{N}a_{i}|m_{i}>$, $m\in\mathbb{Z}, m<\infty$, and $\hat{x}_{5}$ is defined through its exponentiated version, i.e., $\hat{U}|m>=\frac{1}{2}\left(|m+1>-|m-1>\right)$. Because of (\ref{complex}), it is not conceptually difficult to chose a polarization instead of the other, because of Born reciprocity, but the two polarizations bring very different results indeed. In the $p$ polarization, $\psi(P_{5})=\delta_{P_{5},m}$, so that the position space is compactified {\it because} $m\in\mathbb{Z}$, and the wave functions are quasi-periodic functions on the Bohr compactification of $\mathbb{R}$. As a result, the model can be interpreted as compactified on a "fuzzy" circle\cite{corfuz}, because $m<\infty$. Since non-commutative geometries do not apply in 1 dimension, this result can be looked to as an alternative construction of the fuzzy circle, which can be obtained from a fuzzy 2-sphere\cite{dol03}, with similar cut-off properties.
\section{Modified Commutation Relations: Generalized Uncertainty Principle}
It is however possible to modify the canonical commutation relations between the extraD operators $\hat{x}_{5}$ and $\hat{P}_{5}$. Within the framework of a higher number of extraD, compactification on non-commutative spaces has been widely investigated \cite{con}. In 5D, nevertheless, it is obviously impossible to implement non-commutative coordinates; contrastingly, modified commutation relations are the tool that can allow for a similar procedure\cite{kem94,nou07}. Since unitary transformations preserve commutation relations, predictions based on the GUP approach cannot be compared to ordinary ones.\\
The most general commutation relations that can be considered reads\cite{kem94}
\begin{equation}\label{modicom}
[\hat{x}_{5},\hat{P}_{5}]=i\hbar(1+\alpha\hat{x}_{5}^2+\beta\hat{P}_{5}^2).
\end{equation}
Nonetheless, because of the problem we are investigating, it would sound sensible to set $\beta\equiv0$ in (\ref{modicom}): according to Ref.\refcite{kem94}, no momentum eigenstate can be found, since it would have zero uncertainty. Information about momentum can be recovered by investigating ''maximal-momentum-localization'' states, which generalize plane waves in the coordinate representation, and then using the projection on ''quasi- momentum'' wave functions for general states\footnote{So far, this description is dual to that presented in Ref.\refcite{kem94}, but an interpretative mismatch occurs, when the dispersion relations corresponding to this modified scheme are looked for. In fact, although the Fourier duality between position and momentum is widely accepted, the Fourier dual of a dispersion relation is not so clearly understood. Furthermore, the interpretative difficulties of Fourier dualism arising from non-standard geometries\cite{maji00} is due to the fact that Born reciprocity is somehow deformed by (\ref{modicomma}).}. We can however retain the main idea of considering modified commutation relations to investigate the implications of such a model, and adopt the following commutation relations,
\begin{equation}\label{modicomma}
[\hat{x}_{5},\hat{P}_{5}]=i\left(1+\frac{P_{5}^{2}}{N^{2}}\right),
\end{equation}
where the parameter $\alpha$ in \ref{modicom} has been set equal to zero, and $\beta$ has been connected with the maximum number of particles, i.e., $\beta\equiv1/N^2$. In this representation, we find a modified dispersion relation,
\begin{equation}\label{lambda}
\lambda=\frac{2\pi}{N}\frac{1}{\arctan \frac{m}{NL}}.
\end{equation}
If we consider $N$ as an independent parameter of the theory and let $m\rightarrow\infty$, we obtain a minimal wavelength $\lambda_{N}=4/N$, which vanishes for $N\rightarrow\infty$, when canonical commutation relation are recovered. Contrastingly, we can consider $N$ as the maximum number of particles allowed in the model, and consequently, $m_{max}=N$: in this case, the minimal length that characterizes this scheme is $\lambda_{N,L}=(2\pi)/(N\arctan \frac{1}{L})$. In both cases, a minimal length is predicted, and the error in considering $\lambda_{N,L}$ instead of $\lambda_{N}$ can be made negligible even by shrinking the compactification scale\cite{kub93}, $L$, while keeping $N$ large but finite.

\end{document}